# Are Generation Z Less Car-centric Than Millennials?

# A Nationwide Analysis Through the Lens of Youth Licensing

**Kailai Wang, Ph.D.**

Assistant Professor of Supply Chain and Logistics Technology

Department of Construction Management

Cullen College of Engineering, University of Houston

4730 Calhoun Road #332, Houston, TX 77204-4021

# Are Generation Z Less Car-centric Than Millennials?

# A Nationwide Analysis Through the Lens of Youth Licensing

**Abstract**

Are young Americans becoming less auto-centric? According to National Household Travel Survey (NHTS) data, there is a noticeable decrease in the rate of driver's license acquisition among teens aged 16–20 years. 65.4% of this age group members held a driver's license in 2017, which is 8.3% and 8.1% lower than in 2001 and 2009, respectively. This research compares the differences in driver's license acquisition between Millennials (teens in 2009) and their succeeding generation – Generation Z (teens in 2017) during late adolescence. This research also investigates the factors that influence a teen's decision to hold a driver's license. Findings suggest that the dropped licensing rate between two survey years can be explained by a generational shift in attitudes and cultural changes to a certain extent. Some trait and characteristics of Generation Z, such as making more educational trips and growing up in a digital world, may significantly influence their decisions about obtaining a driver's license. To further explore whether Generation Z will drive less than Millennials once getting a driver's license, this research conducts a multivariate analysis for licensed teens in 2009 and 2017, focusing on their driving distances on the survey day. At least, in this research, I do not find a significant difference between teens aged 18-20 years from two generations. This research draws out the implications for planners, practitioners, and policymakers on proactively responding to the possible consequences of changing American car culture.

**Introduction**

Planners and policymakers are interested in knowing how social, cultural, and historical contexts shape people's habits and transportation needs. For instance, the long-lasting effects of the Great Recession from 2007 to 2009 on the "Millennial generation" (those born between 1981 and 1996) has captured considerable attention in the US and many other developed countries, including media outlets, academic articles, and planning actions. Transportation scholars and professionals have attempted to understand the causes of delaying licensure and reducing private auto ownership and usage (Blumenberg et al., 2016; da Silva et al., 2019; Knittel & Murphy, 2019; McDonald, 2015; Wang, 2019; Wang & Akar, 2020; Zhang & Li, 2022). If there is a generational shift in residential and travel preferences (Thigpen & Handy, 2018; Lee et al., 2020), public service sectors and private businesses should proactively seek solutions to accommodate the future population. Nowadays, the demographic cohort succeeding Millennials – Generation Z (or Gen Z for short; those born between 1997 and 2012) – has accounted for about one-fifth of the US population (US Census Bureau, 2020). The generational traits of Gen Z are very likely to shape and profoundly influence our society as they mature and start getting older. Understanding how members of Gen Z behave in their late adolescence has far-reaching implications for long-term transportation policy provisions and infrastructure investments. This study compares the differences in licensure for teens in the US across generations. A particular emphasis has been placed on members of Gen Z and Millennials.

Demographic cohorts are crucial as they can be used to analyze the impacts of historical events and societal changes that occurred at different times (Ryder, 1985). Historical events influenced Millennials and members of Gen Z differently since the two age cohorts were at different stages of their lives. Specifically, most Millennials were old enough to understand 9/11 international terrorism and the 2008 global recession, while members of Gen Z were too young to memorize these historical events. Based on an analysis of the Current Population Survey (CPS) Annual Social and Economic (ASEC) Supplements, the Pew Research Center reported that Gen Z is by far the most racially and ethnically diverse generation in US history (Parker & Igielnik, 2020). They are also the first generation growing up in a truly digital environment with a better educational background than previous generations – so called iGen (Twenge, 2017). Hence, the transportation needs of Gen Z are likely to be different from those of previous generations.

The characteristics of Gen Z and their causes have been initially narrated in media outlets (e.g., Allison+ Partners, 2019; Dimock, 2019; Parker & Igielnik, 2020; Katz et al., 2019). A recent report suggests that members of Gen Z care more about environmental sustainability than previous generations regarding travel mode choice (Allison+ Partners, 2019). For a teen, obtaining a driver's license is a traditional first step toward independence and freedom. It also holds true for today's Gen Z teens. Licensing allows teens to socialize and meet their practical needs (Handy et al., 2021). However, researchers have begun to recognize the decline in youth licensing in North America and other developed countries since around 2010. (Delbosc & Currie, 2013; Delbosc & Currie, 2014; Delbosc, 2017; Habib, 2018; Le Vine et al., 2014; McDonald & Trowbridge, 2009; Rérat, 2021; Thigpen & Handy, 2018; Wu et al., 2021). Youth licensing has clear sustainability implications. The current evidence is sufficient to support the claim that if someone delays holding a driver's license, then he or she is more likely to use private cars less often in later life. Teens who delay licensing may gain valuable skills and traveling experiences by active modes and using public transit (Delbosc, 2017; Smart & Klein, 2018). Admittedly, only time can tell us whether the new mobility patterns of Millennials and Gen Z will persist, a greater amount of empirical research is needed to unravel what determines the observed trend.

**Table 1. Changes in the driver's license acquisition during 2001-2017[1]**

| Age group | 2001 NHTS | 2009 NHTS | 2017 NHTS |
|---|---|---|---|
| 16-20 | 73.67% | 73.48% | 65.38% |
| 21-36 | 91.77% | 90.22% | 88.80% |
| 37-52 | 95.03% | 94.38% | 92.96% |
| 53-71 | 92.05% | 91.97% | 89.72% |
| 72- | 72.92% | 73.30% | 76.82% |
| Overall | 86.19% | 87.51% | 85.76% |

Source: US Department of Transportation's National Household Travel Surveys of 2001, 2009, and 2017.

Table 1 presents the driver's license acquisition trends during 2001-2017 for multiple age groups. A noticeable decline in driver's license acquisition rate among teens aged 16–20 years can be observed. 65.4% of this age group members held a driver's license in 2007, which is 8.3% and 8.1% lower than in 2001 and 2009, respectively. According to the widely accepted definition, members of Gen Z were born between 1997 and 2012, and aged 16-20 years in 2017 (Dimock, 2019). Those teens of the same age in 2009 born between 1989-1993 belong to the Millennials cohort. This study uses the two most recent US National Household Travel Surveys (NHTS) to perform multivariate analyses on travel diary data. The research results reveal the individual, household, and regional level factors associated with a teen's driver's license acquisition. The main contribution of this research is the comparison of the differences between Millennials and members of Gen Z. Using the same correlates of driver's license acquisition, this study further investigates the differences in daily driving distances between members of Gen Z and Millennials, focusing on licensed teens. The findings of this study shed light on whether Gen Z is less auto-centric than previous generations. The findings are expected to aid planners and policymakers in formulating comprehensive provisions to proactively address the unintended consequences of the decline in youth licensing and vehicle ownership reduction.

---

[1] Using personal weights, the study makes the respondents proportionate to the national population and removes biases associated with data collection and non-response.

In the next section, I first review the existing research on Gen Z's behavioral characteristics and youth licensing decline. Then I describe the data, descriptive statistics, and research design. Next, statistical modeling results are presented, followed by a summary of demographic, planning, and policy implications. Finally, I conclude the remarks and suggest pathways for future research.

**Literature Review**

There has been much discussion about generational differences in travel patterns, but less is said about how Gen Z behaves differently. Transportation planning researchers have extensively explored the travel patterns of Millennials with a focus on uncovering the effects of changing travel preferences or attitudes, delayed life course milestones (e.g., education, marriage, parenthood), and macroeconomic events (i.e., the Great Recession from 2007 to 2009) (Blumenberg et al., 2016; da Silva et al., 2019; Garikapati et al., 2016; Knittel & Murphy, 2019; Lee et al., 2020; McDonald, 2015; Wang, 2019; Wang & Akar, 2020; Zhang & Li, 2022). The basic conclusion is that the changing mobility landscape of Millennials is a composite outcome of factors representing all three aspects mentioned above. In recent years, Millennials have started hitting their middle age, and the successive generation – Gen Z is entering adulthood. Considering Gen Z to be the future of our society and its sizeable proportion of the US population, the number of studies on the travel patterns of this generation is disproportionately small. Through the lens of youth licensing, I am taking the first step towards closing this gap.

*The characteristics and travel patterns of Generation Z*

The state-of-art in transportation planning lacks detailed documentation on the traits and characteristics of Gen Z. The oldest members of Gen Z turned 23 years old in 2020 and thereby voted in the 2020 presidential election. In the US and globally, they can have a significant impact on future transportation needs and performance.

Some traits and characteristics of Gen Z have been widely discussed, including their transportation needs (e.g., Allison+ Partners, 2019; Dimock, 2019). The US population has experienced an increase in racial and ethnic diversity since 2010. This is particularly true for Gen Z (Fry & Parker, 2018; Parker & Igielnik, 2020). It is also expected that this generation will be better educated than previous ones (Fry & Parker, 2018). They are passionate about human rights and their identities (Parker & Igielnik, 2020; Rue, 2018). The native use of technology is another characteristic of Gen Z. Millennials were considered digital pioneers witnessed the explosion of social media and technology; Gen Z was born into an era when information was instantly accessible and social media was increasingly ubiquitous (Katz et al., 2021; Twenge, 2017). The overuse of ICTs has adverse effects. As members of Gen Z use ICTs as their primary means of socializing, they become less interested in going out with friends or conducting other social activities than previous generations (Katz et al., 2021; Trinko, 2018). Furthermore, Gen Z witnessed the rise of school shootings, climate change, terrorism, and their parents experienced massive financial hits during the Great Recession (Luttrell & McGrath, 2021). Thus, this generation has become more cautious and pragmatic. Admittedly, the views of Gen Z have not been fully formed and may change as they age, as well as the intervention of national and global events (i.e., COVID-19 pandemic). A glimpse of how these characteristics and experiences of Gen Z's affect the willingness to license is timely and vital.

Researchers in other disciplines have attempted to identify Gen Z's traits and characteristics in different contexts following the generational cohort theory (Mannheim, 1952; Strauss & Howe, 1997). For example, Thach et al. (2020) examined how US Gen Z differs from other age cohorts in wine consumption preferences and behaviors. Cilliers (2017) studied the technology preferences of Gen Z students, such as the ways to receive academic information (i.e., Facebook, WhatsApp, or Not via social media) and exam modes (i.e., electronic or written). From the transportation perspective, Olsson et al. (2020) explored the relationships between public transit use, quality perceptions, and life satisfaction for five generations, including Gen Z. This study found that members of Gen Z ride public transit and travel on foot more often than other age cohorts. Remarkably, public transit usage decrease with age. Olsson et al. (2020) reported that Gen Z is more likely than other age cohorts to travel by car as passengers, which can be explained as this generation views cars as less of a status symbol. In another study, Lee and Circella (2019) suggested that Millennials and Gen Z are more likely to be frequent ICT users and prefer less car-dependent travel choices. However, in both studies, members of Gen Z are less represented. Collecting and analyzing sufficient data using rigorous statistical methods can benefit academic researchers and transportation professionals.

*Factors that influence youth licensing*

Many countries have adopted graduated licensing policies in recent decades to protect teens from encountering high-risk driving situations before they have sufficient driving experience. The programs involve several stages before a driver is given a full license. Each state sets its rules about how young people can get a learner's permit and graduated licenses in the US context. At the age of 18-year-old, teens in all states are eligible for a full unrestricted license (Bates et al., 2018; Hirschberg & Lye, 2020; Williams et al., 2016; Witmer, 2019). Beyond safety benefits, the graduated driver licensing (GDL) laws are also regarded as the reasons for the decline in total travel demand among Millennials during their teen years (Blumenberg et al., 2016). Many aspects of youth licensing have been studied, such as low and declining rates of license acquisition, available options for getting a license, and the wait time.

Based on a sample of US teens aged 15-18 years collected in 2010, Williams (2011) examined the subjective reasons behind driver's licensing delay or cancellation. These reasons include do not have access to a household car, the cost of car ownership and maintenance, living with parents, having other ways to travel, and the amount of time it takes to learn how to drive. Delbosc and Currie (2013) synthesized relevant studies in most developed countries. They argued that youth licensing is primarily influenced by demographic and structural changes. Full-time employment, living independently, and child-rearing increase the likelihood of licensing. Living a car-light lifestyle is also influenced by environmental consciousness and no longer seeing a private car as a status symbol.

The residential built environment and transportation accessibility influence youth licensing substantially (Delbosc & Currie, 2013; McDonald & Trowbridge, 2009; van der Waard et al., 2013). The share of teens with driver's licenses is higher in inner cities or mixed-use and walkable suburbs than small towns and rural areas. Depending on the circumstances, there seems to be a bidirectional relationship between the built environment and youth licensing, which calls for collecting and analyzing longitudinal data.

Focusing on the impacts of demographic trends, Delbosc and Currie (2014) analyzed Australian youth aged 18-30 years using data from four repeated cross-sectional household travel surveys. Unlike previous studies, the modeling results showed that the likelihood of obtaining a driver's license increased when living with parents. This may be owning to increased auto access and the effects of experienced drivers in the household. Brown and Handy (2015) noted that parental encouragement could positively influence youth licensing. However, if some parents often chauffeur their children to daily activities, it tends to dampen youth licensing. After controlling for all other factors, Delbosc and Currie (2014) further pointed out that young adults in 2007 and 2009 were less likely to possess a driver's license than their counterparts in 1994. Hjorthol (2016) examined how car ownership has declined among Norwegian young adults over the last two decades and found that they are becoming less interested in learning to drive. The influential factors are occupation, place of living, use of public transportation, education, and marital status. This trend analysis showed that youth are delaying starting families and finishing their education, resulting in a longer waiting period for their driver's licenses.

The timing of getting a license is another focus of youth licensing literature, which concisely links behavioral insights to policy interventions (Habib, 2018; Tefft et al., 2014; Thigpen & Handy, 2018). A measure of licensure age can be used as a more nuanced indicator of when and how drivers are issued licenses. Teens without a driver's license will likely be forced to use alternative forms of transportation. Habib (2018) surveyed post-secondary students aged 15-35 years from four universities in Toronto and developed the hazard-based duration model to investigate the determinants associated with the decision and age of obtaining a driver's license. The most substantial and negative influential factor is living with parents. Improved transit service quality and transportation accessibility can delay or even forgo the need for driving. Apartment and condo dwellers in the central city are more likely to delay licensure.

Existing studies have discussed the interactions between transportation outcomes and information and communication technologies (ICTs) usage. Some earlier studies suggested that e-communications influence teens' willingness to hold a driver's license as they allow contacting others via social media platforms while traveling by public transit (Delbosc & Currie, 2013). A deeper exploration of online activities to youth licensing can be found in Le Vine et al. (2014). This study suggested that some evidence supports the existence of a nonlinear relationship between the intensity of virtual activity and driver's license acquisition. In other words, substitution and complementarity effects exist under different conditions.

In summary, the decline in youth licensing in developed countries and Gen Z's behavioral characteristics have caught the attention of researchers and professionals. However, little is known about the connection between the two streams of literature. This study contributes to the literature by separating members of Gen Z from Millennials. Some researchers have discussed the travel patterns of Gen Z, together with other age groups (Lee & Circella, 2019; Olsson et al., 2020). Based on the US nationally representative travel surveys, a comprehensive investigation of the differences between Gen Z and Millennials regarding their driver's license acquisition is meaningful from theoretical, behavioral, and policy perspectives.

**Data and Methods**

*Data sources, variables, and descriptive statistics*

This study presents and analyzes information from the three most recent nationwide travel surveys – the 2001, 2009, and 2017 National Household Travel Surveys (NHTS). As shown in Table 1 (in the Introduction section), the 2001 NHTS data is included to observe the changes in the driver's license acquisition over a more extended period. Notably, the NHTS data collections were influenced by major social and historical events. First, the 2001 NHTS sample was collected during March 2001 and May 2002, including the 9/11 international terrorism. As a result of the attacks, transportation services were disrupted for months, particularly curtailing long-distance trips during the winter holidays. The volume and mode choice of urban travel might be influenced to an unknown and probably large extent (Erlbaum, 2005; Pucher & Renne, 2003). Second, the 2009 NHTS was conducted in the midst of a severe economic recession. For multivariate statistical analysis, this study conducts a research dataset using samples from the 2009 and 2017 NHTS. The differences in responses between the two survey years reveal the changing economic and social conditions.

It should also be cautious with survey designs and implementations at different years. In 2001 and 2009, a random-digit-dialing (RDD) landline telephone sampling strategy was adopted. The sample was stratified by geographic features, such as the census divisions, metropolitan area size, and access to heavy rail transit. Personal information was acquired by telephone interviews and computer-assisted telephone interviews (CATI) in 2001 and 2009, respectively. In contrast, the 2017 survey used an address-based sampling approach that includes both households with and without landline telephones (e.g., some households have cellphones only), namely, a web-based scheme (Federal Highway Administration, 2004, 2018). It is likely that the 2009 sample might have underreported those teens who frequently use mobile phones but lack access to landline telephones. Indeed, many teens in 2009 used mobile phones.

The NHTS data contains information on individuals, households, vehicles, and detailed travel diaries on a survey day. The publicly available geographic identifiers are metropolitan statistical area (MSA), state, and census region and census division. Each survey data file provides a sample weight that can inflate the respondents to the population, reducing biases associated with the data collection process. This study adopts the personal sample weights to execute descriptive statistics and two-sample t-tests below. In order to generate nationally representative statistics, survey weights adjust for non-responses. I do not adjust for personal weights in multivariate regression analysis as to the accuracy of standard errors (Winship & Radbill, 1994). This is because some factors (e.g., geographic areas) are directly used in the development of survey weights. I tested models with and without survey weight, and obtained similar results.

Members of Gen Z are those born between 1997 and 2012. The preceding generation Millennials born between 1981 and 1996. (e.g., Dimock, 2019). The former cohort members were 5–20 years old in 2017, and the later ones were 13–28 years old in 2009. In the NHTS datasets, a driver is defined as a person who is at least 16 years old and drives a vehicle on the survey day for at least one trip. Within these considerations, this study builds a research dataset comprising teens between the ages of 16 and 20 from the 2009 and 2017 NHTS. The final study sample excludes some responses that did not provide sufficient information on our variables of interest. Table 2 presents the descriptive statistics for the study sample, which contains 8,009 and 8,543 teens from the 2009 and 2017 NHTS, respectively. The status of driver's license acquisition is the dependent variable in this study. Table 3 reports a substantial decline in driver's license acquisition rate at each age. In addition, I have compared the research sample to teens aged 15-19 from the American Community Survey (ACS) based on gender, employment status, and race and ethnicity groups. The results suggest that this study has a nationally representative sample. According to the Literature Review section, a teen's decision to hold a driver's license can be influenced by seven types of factors: 1) socio-demographic characteristics, 2) life course events, 3) household incomes and vehicle ownership, 4) current travel mode choice (including ICT usage), 5) number of daily trips on the survey day, 6) residential location choice, and 7) census regions and divisions.

  The 2009 NHTS did not collect education attainment data from teens under 18 years old. At the same time, this information is crucial to the possession of the driver's license and reflects the characteristics of Gen Z to a point (Twenge, 2017). Thus, this study focuses on teens aged 18-20 years who reported their educational background in multivariate analysis. Being employed and leaving parents' residence will likely stimulate the travel needs of individuals under 20 years old. Living with parents is associated with being financially reliant, lowering housing costs, and having access to household vehicles. This study uses these two factors to measure life course events. Also, the percentage of family members with a driver's license is calculated. If a teen from a household having other household members, I calculate the percentage of other family members with a driver's license (i.e., excluding this person during the calculation); if a teen does not have any other household members, the variable is set to zero. By doing so, parental and peer effects on getting a license are captured.

**Table 2. Data Summary for research sample (16-20 years old)**

| Survey year | 2009 NHTS Mean/% (±SD) | 2017 NHTS Mean/% (±SD) |
|---|---|---|
| *Driver's license acquisition* | | |
|   Holder | 73.8% | 67.1% |
|   Non-holder | 26.2% | 32.9% |
| *Socio-demographic characteristics* | | |
|   Age | 18.02 (±1.33) | 17.97 (±1.39) |
|   Gender | | |
|     Male | 51.6% | 52.9% |
|     Female | 48.4% | 47.1% |
|   Educational Attainment [a] | | |
|     Less than high school | 21.2% | 46.5% |
|     High school graduate | 41.0% | 28.4% |
|     Some college or associate degree | 36.4% | 24.4% |
|     Bachelor's and higher | 1.4% | 0.7% |
|   Immigration Status | | |
|     Born in US | 90.5% | 91.5% |
|     Foreign-born | 9.5% | 8.5% |
|   Race of household head | | |
|     Non-Hispanic white | 62.2% | 53.9% |
|     Non-Hispanic black | 13.6% | 13.7% |
|     Non-Hispanic Asian/Pacific Islander | 3.1% | 4.9% |
|     Hispanic | 18.8% | 20.9% |
|     Other races | 2.3% | 6.6% |
| *Life course events* | | |
|   Employment status | | |
|     Worker | 49.7% | 45.7% |
|     Non-worker | 50.3% | 54.3% |
|   Formed household [b] | | |
|     Lives independently | 2.4% | 2.1% |
|     Lives with parents or other older relatives | 97.6% | 97.9% |
| *Household incomes and vehicle ownership* | | |
|   Percent of family members with a driver's license [c] | 0.68 (±0.29) | 0.68 (±0.28) |
|   Household vehicle ownership | | |
|     Zero-vehicle household | 4.9% | 4.4% |
|     One-vehicle household | 16.7% | 15.4% |
|     Two-vehicle household | 25.1% | 23.0% |
|     Households with three or more vehicles | 53.3% | 57.2% |
|   Household incomes per capita (in 1,000 2017 US$) [d] | 35.88 (±25.96) | 43.18 (±34.31) |
| *Current travel mode choice* | | |
|   Number of walk trips in the past week | 4.66 (±6.48) | 5.55 (±9.54) |
|   Number of bike trips in the past week | 0.43 (±2.44) | 0.36 (±1.87) |
|   Number of days used public transit in the past month | 3.67 (±8.02) | 2.33 (±6.19) |
|   Frequency of purchased online for delivery in the past month | 0.57 (±1.51) | 1.42 (±3.07) |
|   Usage frequency of rideshare services (e.g., Uber and Lyft) in the past month | | 0.37 (±2.08) |
|   Usage frequency of carshare services (e.g., Zipcar and Car2go) in the past month | | 0.01 (±0.35) |
| *Number of daily trips on the survey day* | | |
|   Social trips | 0.63 (±0.91) | 0.52 (±0.84) |
|   Recreation trips | 0.24 (±0.57) | 0.15 (±0.41) |
|   Exercise trips | 0.18 (±0.47) | 0.10 (±0.34) |
|   Errand trips | 0.80 (±1.16) | 0.59 (±1.06) |
| Education trips | 0.38 (±0.62) | 0.45 (±0.61) |
| Work trips | 0.34 (±0.63) | 0.33 (±0.70) |
| *Residential location choice* | | |
| Population density (1,000 persons per square mile) [e] | 6.16 (±7.49) | 4.88 (±6.60) |
| Size of metropolitan areas | | |
|   Not residing in MSAs | 12.1% | 15.5% |
|   In an MSA or CMSA of population less than 1,000,000 | 22.8% | 32.2% |
|   In an MSA or CMSA of 1,000,000 - 2,999,999 | 22.7% | 17.7% |
|   In an MSA or CMSA of 3 million or more | 42.4% | 34.6% |
| MSA heavy rail status for household | | |
|   MSA has rail | 34.6% | 26.2% |
|   MSA does not have rail, or household not in an MSA | 65.4% | 73.8% |
| Urban / Rural indicator | | |
|   Urban | 21.6% | 15.6% |
|   Suburban | 28.6% | 23.7% |
|   Second City | 19.5% | 20.1% |
|   Small Town and Rural | 30.3% | 40.7% |
| *Census regions and divisions* | | |
|   New England | 4.9% | 4.0% |
|   Middle Atlantic | 15.2% | 12.5% |
|   East North Central | 14.5% | 15.8% |
|   West North Central | 4.8% | 7.0% |
|   South Atlantic | 16.6% | 18.5% |
|   East South Central | 4.8% | 5.9% |
|   West South Central | 11.0% | 13.5% |
|   Mountain | 7.9% | 7.4% |
|   Pacific | 20.4% | 15.5% |
| Number of observations | 8009 | 8543 |

Notes: (a) – Respondents less than 18 years old did not provide education attainment information in 2009 NHTS. The number of respondents having completed education information in the final sample is 3,802 and 4,251, respectively; (b) – Lives independently means do not have elder family members; (c) – If a teen from a household having other household members, I calculate the percentage of family members with a driver's license; if a teen does not have any other household members, the variable is set to zero; (d): The midpoints of the categories are used in this study. Whenever a respondent reports the annual household incomes in the range of $10,000 to $14,999, the value is taken as $12,500. For the 2009 NHTS, I set the highest interval of "$100,000" as "$150,000"; in 2017 NHTS, the highest interval of "$200,000" equals "$250,000". As the purchasing power fluctuates as time goes on, I therefore convert the incomes of 2009 into 2017 US dollars (https://www.usinflationcalculator.com/). To measure household income per capita, I finally use square root scaling (http://www.oecd.org/els/soc/OECD-Note-EquivalenceScales.pdf); (e) – Similar to the household incomes per capita, the midpoint of each category is utilized to construct a continuous indicator of population density. I use 30,000 persons per square mile for the most densely populated category, in terms of ⩾25,000 people per square mile).

**Table 3. Driver's license acquisition rates by age**

| Age | 2009 | 2017 | Difference |
|---|---|---|---|
| 16 | 48.6% | 43.1% | -5.5% |
| 17 | 68.8% | 62.9% | -5.9% |
| 18 | 78.5% | 71.3% | -7.3% |
| 19 | 84.2% | 79.8% | -4.4% |
| 20 | 84.8% | 79.4% | -5.4% |

As shown in Table 2, there are some similarities between Millennials and members of Gen Z, but also some noteworthy differences. Among teens aged 16-20 years, there was an 8.3% decline in the white share of the total population during 2009-2017. The data confirms that the US population is becoming more racially and ethnically diverse. The values of household income per capita suggest that members of Gen Z grew up in a more affluent condition than their Millennial counterparts. This study includes a list of explanatory variables for measuring residential location choice and geographic effects – residential density, size of metropolitan area (MSA), MSA heavy rail status, urban and rural indicator, and census regions and divisions. These variables can serve as proxies for the level of compact development, household residential preferences, and transport accessibility. Combined with the differences in household income per capita and residential location choice between two cohorts, it is not difficult to discern whether the social and economic context has changed during 2009-2017. The recovery from the Great Recession increases US household incomes and makes more teens in 2017 reside in lower-density neighborhoods with fewer urban amenities than their counterparts in 2009.

Figure 1 provides a state-level observation of the driving license acquisition changes during 2009-2017. Issues such as gas tax, political orientation, and licensing requirements influence a teen's attitudes towards driving. The decline in licensure occurred in more than 40 US states. Kansas, Mississippi, Oregon, and the District of Columbia experienced a more than 30% decline in the rate of youth licensing. To better understand the car culture shift in a spatial domain, Table 4 reports the statistical summaries of driver's license rates among teens aged 16–20 years for seven regions. Teens living in the western US, namely Pacific and Mountain regions, experienced a significant drop in driver's license acquisition. While I do not observe a significant decline in youth licensing in the Southern regions of the US, which reflects that "getting a driver's license" is more deeply rooted in these regions.

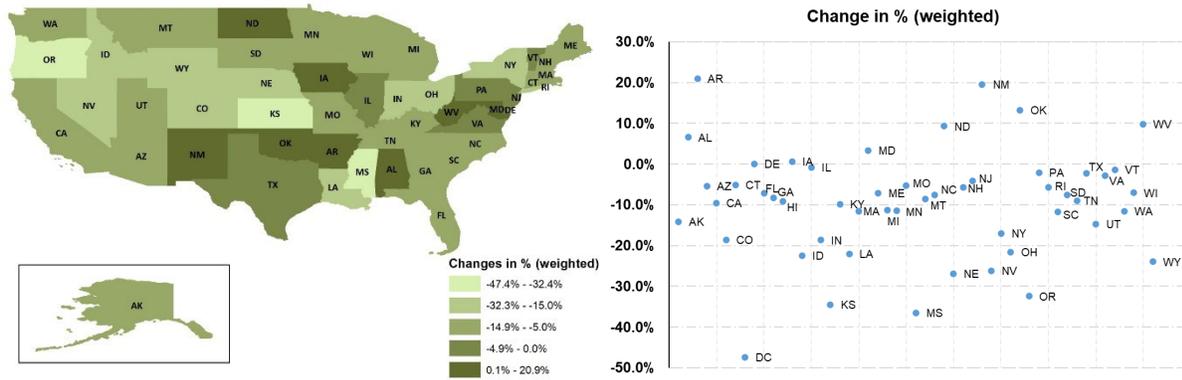

Figure 1. Changes in the state-level driver's license acquisition during 2009-2017

Table 4. Driver's license acquisition across census divisions

|  | 2009 | | | 2017 | | | 2009 versus 2017 | |
| --- | --- | --- | --- | --- | --- | --- | --- | --- |
|  | Obs. | Mean | SD | Obs. | Mean | SD | Changes in % | *p* value |
| Census Division [a] | | | | | | | | |
| Northeast | 1274 | 0.67 | 0.47 | 1304 | 0.59 | 0.49 | -11.6% | 0.092 |
| East North Central | 438 | 0.78 | 0.41 | 990 | 0.69 | 0.46 | -11.5% | 0.118 |
| West North Central | 458 | 0.83 | 0.38 | 324 | 0.71 | 0.46 | -15.0% | 0.089 |
| South Atlantic | 2248 | 0.76 | 0.42 | 1850 | 0.72 | 0.45 | -5.2% | 0.199 |
| South Central | 1228 | 0.70 | 0.46 | 2078 | 0.73 | 0.45 | 4.0% | 0.522 |
| Mountain | 506 | 0.89 | 0.31 | 312 | 0.68 | 0.47 | -24.4% | 0.000 |
| Pacific | 1857 | 0.70 | 0.46 | 1685 | 0.58 | 0.49 | -17.1% | 0.000 |

Notes: (a) – The original census divisions are classified into seven categories considering the spatial proximity and transportation characteristics – Northeast (i.e., the combined area of New England and Middle Atlantic divisions), South Atlantic, South Central (i.e., combined area of East South Central and West South Central divisions), East North Central, West North Central, Mountain, and Pacific.

*Changes in teens' travel mode choice and daily activities*

This study hypothesizes that teens in 2017 should generate more trips than their counterparts in 2009 due to the economic context. Consistent with Shirgaokar and Nobler (2021), I categorize daily trips into six groups. The data in Table 5 shows that teens aged 16-20 in 2017 travel more for non-discretionary needs (working and educational purposes), regardless of the status of licensed or not. The data also tells us that both driver's license holders and non-holders of the studied age group participated in discretionary activities less frequently in 2017, relative to their counterparts in 2009. Previous studies report that youth travel less in 2009 are mainly due to heavy financial burden and lack of income. The continued decline in 2017 indicates that the changes in daily activity and travel patterns across generations. Perhaps, members of Gen Z are even more "go-nowhere" than Millennials.

This study measures current travel mode choice from the following aspects: walking and bicycling frequencies, online shopping frequency, and usage frequency of on-demand mobility services. Table 6 displays the differences in travel mode choice between two generations for both driver's license holders and non-holders. Teens in 2017 generated more waking trips than their counterparts in 2009. The result shows the opposite when it comes to bicycling trips. Teens in 2017 may use bikeshare programs to substitute traditional bicycle trips. Note that active travel can replace short auto trips and reflect one's attitudes towards cars. Public transit ridership dropped significantly during 2009-2017 among licensed teens. For non-licensed teens, the frequencies of public transit ridership in 2009 and 2017 are almost the same. Teens in 2017 purchased online much more often than their counterparts in 2009. This is not surprising as Gen Z is the first to be described as "digital natives".

**Table 5. Number of trips generated on the survey day by teens in 2009 and 2017**

|  | 2009 | | 2017 | | 2009 versus 2017 | |
| --- | --- | --- | --- | --- | --- | --- |
|  | Mean | SD | Mean | SD | Changes in % | p value |
| Driver's License Holders | | | | | | |
|    Social trips | 0.67 | 0.94 | 0.57 | 0.87 | -15.3% | 0.006 |
|    Recreation trips | 0.27 | 0.61 | 0.16 | 0.45 | -40.8% | 0.000 |
|    Exercise trips | 0.18 | 0.47 | 0.11 | 0.35 | -38.1% | 0.000 |
|    Errand trips | 0.79 | 1.19 | 0.59 | 1.09 | -24.7% | 0.000 |
|    Education trips | 0.46 | 0.64 | 0.51 | 0.66 | 9.7% | 0.004 |
|    Work trips | 0.33 | 0.61 | 0.37 | 0.65 | 12.5% | 0.795 |
|    Total trips | 2.70 | 1.78 | 2.31 | 1.67 | -14.5% | 0.000 |
| Number of observations | 6112 | | 6107 | | | |
| Driver's License Non-holders | | | | | | |
|    Social trips | 0.57 | 0.88 | 0.45 | 0.81 | -20.6% | 0.024 |
|    Recreation trips | 0.22 | 0.56 | 0.14 | 0.42 | -36.3% | 0.011 |
|    Exercise trips | 0.15 | 0.41 | 0.08 | 0.29 | -46.3% | 0.030 |
|    Errand trips | 0.63 | 1.03 | 0.56 | 0.99 | -11.2% | 0.585 |
|    Education trips | 0.53 | 0.64 | 0.58 | 0.60 | 10.5% | 0.193 |
|    Work trips | 0.11 | 0.39 | 0.14 | 0.37 | 24.2% | 0.098 |
|    Total trips | 2.21 | 1.48 | 1.95 | 1.47 | -11.5% | 0.041 |
| Number of observations | 1897 | | 2436 | | | |

**Table 6. Teens' current travel mode choice in 2009 and 2017**

|  | 2009 | | 2017 | | 2009 versus 2017 | |
| --- | --- | --- | --- | --- | --- | --- |
|  | Mean | SD | Mean | SD | Changes in % | p value |
| Driver's License Holders | | | | | | |
|    Number of walk trips in the past week | 4.22 | 6.33 | 4.62 | 7.74 | 9.5% | 0.136 |
|    Number of bike trips in the past week | 0.35 | 1.45 | 0.25 | 1.29 | -28.2% | 0.057 |
|    Number of days used public transit in the past month | 3.31 | 7.72 | 1.14 | 4.17 | -65.6% | 0.000 |
|    Frequency of purchased online for delivery in the past month | 0.67 | 1.67 | 1.63 | 2.87 | 142.4% | 0.000 |
| Number of observations | 6112 | | 6107 | | | |
| Driver's License Non-holders | | | | | | |
|    Number of walk trips in the past week | 5.89 | 6.74 | 7.43 | 12.22 | 26.1% | 0.039 |
|    Number of bike trips in the past week | 0.64 | 4.09 | 0.60 | 2.69 | -6.9% | 0.894 |
|    Number of days used public transit in the past month | 4.71 | 8.73 | 4.75 | 8.50 | 0.9% | 0.951 |
|    Frequency of purchased online for delivery in the past month | 0.28 | 0.84 | 0.98 | 3.38 | 253.7% | 0.000 |
| Number of observations | 1897 | | 2436 | | | |

*Multivariate statistical methods*

To understand which factors can influence Millennials and Gen Z differently, this study estimates logit regression models that predict the likelihood of acquiring a driver's license. I restrict the data sample to teens aged 18 or older for two reasons. First, for those teens younger than 18 years old, the 2009 NHTS did not document their highest level of education completed. Second, state licensing standards vary considerably, and teenagers in some states cannot obtain a full license before 18 years old.

The issue of multicollinearity was checked for all explanatory variables listed in Table 2 before running statistical models (Freedman, 1991). Theoretically, personal income or household wealth may have a substantial influence on auto ownership. Access to auto is also likely to be highly endogenous with licensing. The urban/rural indicator (i.e., urban, suburban, second city, and small town and rural) in the NHTS datasets is mainly derived from population density at the block group level. Therefore, variables representing household vehicle ownership and urban/rural indicator do not include in multivariate analyses. In addition, census regions and divisions are excluded as this study includes "state" dummy variables to control the effects of state-varying factors (e.g., gas tax and state-level regulations). In Table 7, Models 1 and 2 are estimated for teens aged 18-20 years in 2009 and 2017, respectively. Model 3 is based on the pooled dataset of the 2009 and 2017 samples, which provide a baseline for a further investigation on the differences between Millennials and Gen Z.

The estimated parameters of these logit models cannot be intuitively interpreted as linear marginal effects. To improve the interpretability of regression coefficients, this study computes average marginal effects (AMEs) and average elasticity effects (AEEs) of categorical and continuous variables, respectively. A marginal effect refers to the changes in the percentage of the outcome variable due to a one-unit adjustment of an input variable while holding the rest constant. The elasticity effect corresponds to the percentage change in the outcome variable as a result of a one-percent change in an input variable, *ceteris paribus*. Next, this study explores the differences in daily driving distances between Millennials and members of Gen Z who hold a valid driver's license. As the outcome variable in this step representing driving distance is left-censored, I estimate Tobit models in Table A2 (in the Appendix).

**Results and Discussion**

*Differences in the correlates of driver's license acquisition*

This section begins by discussing modeling results reported in Table 7. The pooled model suggests that, after controlling for other explanatory variables, members of Gen Z have a lower likelihood of holding a valid driver's license at the ages of 18 and 20 as compared to Millennials. The estimated margin shows that the probability of a teen in 2017 obtaining a driver's license is 5.24% less than that of his or her counterpart in 2009. Considering that nearly 80% of teens aged 18-20 years are license holders, the estimated difference is non-trivial. The reason for this could be people's attitudes towards cars are changing across generations.

As shown in Gen Z and pooled models, it is more common for men to hold a driver's license than women. This study does not observe the significant gender difference in obtaining a driver's license among Millennials when other conditions are controlled. Somewhat surprisingly, teens aged 19-20 years are less likely to obtain a driver's license as compared to their counterparts aged 18 years in both Gen Z and pooled models. The result should be verified by analyzing data collected from a broad range of ages. Teens with higher education levels are more likely to obtain a driver's license. Due to the sample size issue, in Gen Z model I do not observe a significant difference between teens completed a Bachelor's and higher degree and those less than high school. This study does not find any significant difference between US-born teens and their immigrant counterparts for both age cohorts. In Millennials model, it shows that non-Hispanic blacks are less likely to hold a driver's license than non-Hispanic whites. Gen Z teens with a white household head are not significantly different from other ethnic groups regarding driver's license acquisition. The results suggest that the increasing diversity of the US population might not have a direct association with youth licensing.

As expected, teens who worked for pay in the past week are more likely to possess a driver's license than those who are unemployed. Individuals who leave their parents' home or live independently have a greater likelihood of getting a license. This is not surprising since holding a driver's license allows a teen to do car-related activities, which will make his or her life easier. Percent of family members with a driver's license is statistically significant and positively associated with the likelihood of obtaining a driver's license. The estimated margins show that these two factors have much stronger correlations with a teen's likelihood of acquiring a driver's license, relative to other explanatory variables in our models. The estimated margins further reveal that, as compared to Millennials, there is a stronger correlation between these two factors and the likelihood of having a driver's license among member of Gen Z. The result implies that Gen Z might have a closer relationship with their parents. Household incomes per capita is positively associated with the outcome variable in our models.

Being a bicyclist, public transit user, or Transportation Network Company (TNC) user is statistically significant and negatively associated with the likelihood of acquiring a driver's license. This can be because some teens with a driver's license could make auto trips for some daily activities. Teens who shop online more frequently are more likely to hold a driver's license as compared with those who buy online less often, which holds true for both Millennials and members of Gen Z. The results indicate that living and breathing the digital environment might have a positive effect on Gen Z's willingness to obtain a driver's license. The effects of new technology-facilitated activities (i.e., TNCs services and e-shopping) on driver's license acquisition should be complex and context-dependent. This finding is line with the results reported by Le Vine et al., (2014).

Teens who travel more and participate in out-of-home activities frequently are more likely to have a driver's license. Both the number of social and recreational trips made on the survey day are statistically significant and positive predictors in Millennials model. However, they are not significant predictors of having a driver's license in Gen Z model. In other words, Millennials who participate in discretionary activities more often tend to have a greater likelihood of having a driver's license, but this does not hold true for Gen Z. Both generations are more likely to acquire a driver's license if they make more trips for run errands. Interestingly, the number of education trips has a statistically significant and positive effect in Gen Z model. However, it does not perform as an effective predictor in Millennials model. This point calls for more policy attention since it has been frequently reported that members of Gen Z are climbing a longer academic ladder than previous generations.

Population density at residences could represent the level of compact development. This study finds that population density has a statistically significant and negative association with Gen Z's willingness to acquire a driver's license. This is expected since urban elements increase teens' access to more opportunities that do not depend on driving automobiles. This study also explores the link between the sizes of metropolitan statistical areas (MSAs) and driver's license acquisition. I find that the likelihood of having a driver's license for teens who reside in smaller MSAs or out of MSAs is not significantly different from that for their counterparts living in medium and larger MSAs. In addition, access to heavy rail system is not a significant predictor in all three models. People who live in larger MSAs are more likely to face the trade-off between housing costs and travel expenses. The results of this study can partially reveal that such a trade-off does not have a direct association with a teen's willingness to obtain a driver's license.

*Cross-generational comparison of driver's license holders*

Following the above discussion, an extended question may be raised: how do these factors affect licensed teens' automobile usage of two generations? Table A1 (in the Appendix) shows a comparison of licensed teens aged 18-20 in 2009 and 2017 for the driving distance and total number of trips made on the survey day. The results of the weighted two-sample t-test suggest that Gen Z's license holders in 2017 drove longer distances than their Millennial counterparts in 2009; however, the difference is not statistically significant. It is notable that although an 11.9% increment in driving distance occurred to licensed teens, the number of total trips made by members of this group has experienced a significant decline during 2009-2017. Combined with the findings from Table 5, I infer that Gen Z teens travel less than teens from the preceding generation. Concisely, they participate in virtual activities more often. Regarding the daily driving distance, a multivariate regression model is considered to control confounding effects.

     In Table A2 (in the Appendix), the dummy variable "year 2017" represents the survey year, indicating whether teens aged 18-20 in 2009 and 2017 performed differently in daily driving distances. Consistent with the results of bivariate statistics, the coefficient of the year dummy variable is not statistically significant in the Tobit model. Other major findings could be summarized as follows. Living independently and the percentage of family members with a driver's license are the two most influential factors to a teen's driving license acquisition, but they are not effective predictors of a licensed teen's driving distance. Immigrant status is not significantly associated with youth licensing; however, the Tobit model's estimates reveal that

native-born teens drive longer distances than their immigrant counterparts among licensed teens. This is not surprising as most existing literature has documented immigrants are less reliant on automobiles than native-born residents, particularly during the early years when they first arrived in the US (e.g., Lee et al., 2021). Perhaps surprisingly to some, this point does not hold for licensed Gen Z teens. As a proxy for virtual activities, online shopping frequency is not a significant predictor of licensed teens' daily driving distances. Similar to the effects of daily activity-travel patterns in driver's license acquisition models, I observe that for both generations, if a licensed teen makes more trips on the survey, then he or she will have a longer daily driving distance.

**Table 7. Logit models of acquiring a driving license (18-20 years old)**

| | Model 1 – 2009 | | | Model 2 – 2017 | | | Model 3 – Pooled Model | | |
|---|---|---|---|---|---|---|---|---|---|
| | Coef. | *p* value | Margins | Coef. | *p* value | Margins | Coef. | *p* value | Margins |
| **Survey year** | | | | | | | | | |
| 2017 (ref: 2009) | | | | | | | -0.618 | 0.000 | -5.24% |
| **Socio-demographic characteristics** | | | | | | | | | |
| Male (ref: female) | 0.063 | 0.616 | 0.49% | 0.235 | 0.029 | 2.10% | 0.169 | 0.034 | 1.44% |
| Age (ref: 18-year-old) | | | | | | | | | |
|    19-year-old | -0.113 | 0.487 | -0.88% | -0.306 | 0.026 | -2.68% | -0.220 | 0.034 | -1.86% |
|    20-year-old | 0.062 | 0.743 | 0.48% | -0.362 | 0.022 | -3.21% | -0.201 | 0.092 | -1.70% |
| Educational Attainment (ref: less than high school) | | | | | | | | | |
|    High school graduate | 0.485 | 0.001 | 3.78% | 0.762 | 0.000 | 6.74% | 0.620 | 0.000 | 5.21% |
|    Some college or associate degree | 1.306 | 0.000 | 10.20% | 1.587 | 0.000 | 13.71% | 1.467 | 0.000 | 11.84% |
|    Bachelor's and higher | 1.574 | 0.092 | 12.29% | 0.306 | 0.577 | 2.60% | 0.910 | 0.046 | 6.51% |
| Born in US (ref: immigrants) | 0.184 | 0.393 | 1.44% | 0.051 | 0.809 | 0.46% | 0.091 | 0.545 | 0.78% |
| Race of household head (ref: non-Hispanic white) | | | | | | | | | |
|    Hispanic | 0.050 | 0.783 | 0.39% | -0.008 | 0.962 | -0.07% | 0.061 | 0.601 | 0.52% |
|    Non-Hispanic black | -0.561 | 0.009 | -4.38% | -0.195 | 0.282 | -1.80% | -0.337 | 0.013 | -3.03% |
|    Non-Hispanic Asian/Pacific Islander | 0.424 | 0.243 | 3.31% | 0.060 | 0.803 | 0.53% | 0.164 | 0.395 | 1.36% |
|    Other races | 0.124 | 0.678 | 0.96% | 0.180 | 0.454 | 1.56% | 0.123 | 0.501 | 1.02% |
| **Life course events, household incomes, and vehicle ownership** | | | | | | | | | |
| Workers (ref: nonworker) | 0.993 | 0.000 | 7.75% | 0.870 | 0.000 | 8.05% | 0.922 | 0.000 | 8.14% |
| Living independently (ref: living with parents) | 1.954 | 0.000 | 15.26% | 3.215 | 0.000 | 16.50% | 2.649 | 0.000 | 13.69% |
| Percent of family members with a driver's license | 5.111 | 0.000 | 22.18% | 5.228 | 0.000 | 27.18% | 5.115 | 0.000 | 24.80% |
| Household incomes per capita (in 1,000 2017 US$) | 0.006 | 0.047 | 1.40% | 0.005 | 0.008 | 1.77% | 0.006 | 0.000 | 1.68% |
| **Current travel mode choice** | | | | | | | | | |
| Number of walk trips in the past week | -0.010 | 0.249 | -0.39% | -0.009 | 0.208 | -0.45% | -0.010 | 0.079 | -0.44% |
| Number of days used public transit in the past month | -0.010 | 0.269 | -0.29% | -0.057 | 0.000 | -1.14% | -0.030 | 0.000 | -0.77% |
| Frequency of purchased online for delivery in the past month | 0.207 | 0.001 | 0.60% | 0.072 | 0.003 | 0.86% | 0.083 | 0.000 | 0.65% |
| Current bicyclists (people who biked in the past week; ref: non-user) | -0.275 | 0.135 | -2.15% | -0.676 | 0.000 | -6.65% | -0.456 | 0.000 | -4.15% |
| User of rideshare services (e.g., Uber and Lyft) in the past month (ref: non-user) | | | | -0.326 | 0.095 | -3.05% | -0.520 | 0.005 | -4.81% |
| User of carshare services (e.g., Zipcar and Car2go) in the past month (ref: non-user) | | | | 0.553 | 0.330 | 4.49% | 0.487 | 0.364 | 3.78% |
| **Number of daily trips on the survey day** | | | | | | | | | |
| Social trips | 0.131 | 0.059 | 0.58% | -0.012 | 0.857 | -0.06% | 0.051 | 0.273 | 0.25% |
| Recreation trips | 0.221 | 0.042 | 0.41% | 0.071 | 0.578 | 0.08% | 0.155 | 0.056 | 0.23% |
| Exercise trips | 0.137 | 0.400 | 0.15% | 0.410 | 0.052 | 0.26% | 0.237 | 0.061 | 0.20% |
| Errand trips | 0.293 | 0.000 | 1.60% | 0.179 | 0.002 | 0.89% | 0.231 | 0.000 | 1.21% |
| Education trips | | 0.144 | 0.201 | 0.42% | 0.240 | 0.015 | 0.87% | 0.204 | 0.006 | 0.68% |
| Work trips | | 0.273 | 0.019 | 0.58% | 0.148 | 0.163 | 0.46% | 0.226 | 0.004 | 0.60% |
| **Residential location choice** | | | | | | | | | | |
| Population density (1,000 persons per square mile) | | -0.004 | 0.745 | -0.19% | -0.027 | 0.011 | -1.14% | -0.015 | 0.051 | -0.69% |
| Size of metropolitan areas (ref: In an MSA or CMSA of 1,000,000 - 2,999,999 & In an MSA or CMSA of 3 million or more) | | | | | | | | | | |
|    Not residing in MSAs | | 0.041 | 0.858 | 0.32% | 0.191 | 0.331 | 1.67% | 0.152 | 0.297 | 1.27% |
|    In an MSA or CMSA of population less than 1,000,000 | | 0.030 | 0.856 | 0.24% | 0.129 | 0.390 | 1.14% | 0.097 | 0.370 | 0.82% |
| MSA heavy rail status for household (ref: MSA does not have rail, or household not in an MSA) | | | | | | | | | | |
|    MSA has rail | | -0.068 | 0.712 | -0.53% | 0.185 | 0.339 | 1.62% | 0.084 | 0.524 | 0.70% |
| Constant | | -2.350 | 0.011 | | -3.184 | 0.000 | | -2.564 | 0.000 | |
| **Model fitness** | | | | | | | | | | |
| Initial log likelihood | | | -1505.46 | | | -1991.53 | | | -3525.99 | |
| Final log likelihood | | | -962.11 | | | -1225.78 | | | -2237.00 | |
| Degree of freedoms | | | 68 | | | 74 | | | 79 | |
| McFadden's R2 | | | 0.361 | | | 0.385 | | | 0.366 | |
| AIC/BIC | | | 2060.22/2483.34 | | | 2599.55/3068.89 | | | 4632.01/5184.01 | |
| Number of observations | | | 3723 | | | 4198 | | | 8002 | |

Note: Due to space constraints, we do not report the coefficients for the state control variables; these are available from the authors upon request.

MSA = metropolitan statistical area; CMSA = consolidated metropolitan statistical area; AIC = Akaike information criterion; BIC = Bayesian information criterion.

**Demographic, Planning, and Policy Implications**

Generational cohort theory is alive and well adopted in the field of transportation planning. As noted by Inglehart (1997), a nation's culture will be shaped by the values of one cohort if they become the majority in the country. Understanding the generations of society is imperative since these values will determine society's values. As of today, Gen Z is growing up to have a driver's license, and their behavior and preferences will influence American car culture. This study reveals many statistical differences between Millennials and Gen Z about driver's license acquisition during late adolescence. The Millennial generation has been observed to be less auto-centric than previous generations in most developed countries. So, will this trend continue and will Gen Z drive less than Millennials? With a detailed analysis of driver's license acquisition, this article provides a preliminary understanding of the factors associated with this question. Below, I summarize the implications of the findings for urban planners and policymakers, as well as future research.

In 2009, Millennials were facing major economic disruptions during their adulthood due to the Great Recession. Members of Gen Z were growing up and living under different social and economic contexts at the same stage of life. The NHTS data suggests that, in relation to Millennials, a lower percentage of members of Gen Z are residing in high-density and central city neighborhoods, and large metropolitan areas. Previous studies have shown great interest in discussing whether Millennials will move out of urban areas when getting a job, starting a family, and having children. This study sheds light on this debate from a different angle. The "coming back downtown" trend (Lee, 2020) does not last for the demographic cohort succeeding Millennials. Choosing urban lifestyles during adulthood is not simply associated with residential location preferences and can also be attributed to economic conditions. Furthermore, a more than 10% increase from 2009 to 2017 in the share of teens aged 16-20 years living in small town and rural areas may merit particular attention. The transportation needs of small town and rural communities will be different from those of urban neighborhoods. Planners and policymakers should consider the ways of creating active travel supportive environments and promoting the quality of public transit services in these communities that can have long-term impacts on teens' travel patterns and subjective well-being in later life stages.

Teens who reside in the western US obtained driver's licenses at a substantially lower rate in 2017 than their counterparts in 2009. To a certain extent, this can be a result of, compared to the rest of the country, West Coast cities and transit agencies made more planning efforts on promoting mixed-use and pedestrian- and transit-oriented development in the past few decades (Jamme et al., 2019). Given the differences between the two survey years, the finding also reemphasizes that changing social attitudes and the adoption of new mobility services may play a role in the decline of teen drivers. In the next step, a useful extension is to collect stated-preference data for a deeper understanding of how geographically distinct cultures and attitudinal factors influence teens' willingness to pay for a vehicle, gas, insurance, maintenance, and other costs exactly.

The trend of increasing racial and ethnic diversity among the US population and its possible effects on transportation outcomes have been extensively discussed in the planning literature (e.g., Blumenberg & Smart, 2014; Chatman & Klein, 2013; Shin, 2017; Lee et al., 2021). Immigrants and minorities use automobiles less than US-born individuals. The findings of this study, however, suggest that this trend is not likely to have a clear connection with Gen Z's driver's license acquisition during late adolescence and their daily driving distances after becoming licensed. Perhaps, planning practices and policy provisions aimed to help immigrants and minorities integrate into the mainstream society have worked effectively; thus, Gen Z is different from other previous generations, at least in the late adolescence. As Smart and Klein (2018) argued that past experiences shape future travel decisions significantly, planners and professionals should be cautious with the revealed generation differences for future travel demand forecasts. Pending the availability of longitudinal datasets, a pertinent direction for future research is to uncover the structural relationships between immigration status, race/ethnicity groups, and automobility focusing on members of Gen Z.

Living independently and percent of family members with a driver's license are much stronger predictors of a teen's driver's license acquisition than other explanatory variables. The estimated margins suggest that two factors are more important to Gen Z than the preceding generation. If the policy goals were to promote travel multimodality and reduce car usage, planners and policymakers should be cautious about the effects of intra-household social relations on teens' travel patterns. It would be helpful if more contextualized and theory-based qualitative research could be conducted to investigate specific mechanisms responsible for household effects on youth licensing.

As expected, teens who travel more often on survey day are more likely to have a license. An earlier study found that, in comparison to the preceding generation, the Millennial generation tends to drive longer distances for social and recreational activities (Wang & Akar, 2020). The study shows that, for driver's license acquisition, Gen Z does not experience a significant effect of participating in discretionary activities. While the number of education trips has a statistically significant and positive relationship with the licensing decision made by members of Gen Z. This point calls for more policy attention as this new generation is more likely to pursue advanced degrees compared to earlier generations.

Results based on the descriptive statistics in Table 5, I infer that these members of Gen Z make fewer trips than Millennials, particularly for non-discretionary needs. It is likely that today's teens have been spending more time online. The popularity of teleworking and e-learning leads them to make fewer trips for working and educational purposes. Relatedly, among licensed teens, this study does not find a significant relationship between the frequency of purchasing online and daily driving distances. The findings suggest that participation in virtual activities can partially substitute for some trips made by teens. However, its influence on daily driving distances or vehicle miles travelled (VMT) remains under-studied. Last but not the least, among license holders, this study does not find any significant differences in daily driving distances between Millennials and members of Gen Z.

**Conclusion**

This study contributes to the literature on generational differences in travel behavior by making a systematic inquiry into how Millennials and the succeeding generation, namely, Gen Z perform differently at the age of 16-20 years about driver's license acquisition. The empirical analyses are based on the data from two most recent National Household Travel Survey (NHTS). The descriptive statistics show that the differences in socioeconomic status and residential location choice between teens in 2009 and 2017 can be attributes to the changing social and economic contexts. Through the estimated logit models, I confirm that the overall drop in the rate of youth licensing during 2009-2017 is a combined result of several individual, household, and regional level factors. The trait and characteristics of Gen Z as well as socio-cultural trends can partially explain why teens in 2017 show less interest in obtaining a driver's license. This study also helps to answer the question, once getting a driver's license, whether members of Gen Z will drive fewer miles in daily life than Millennials. At least, this analysis shows no significant difference between teens aged 18-20 years from two generations. Although we are not able to guarantee and track the travel patterns of Millennials and Gen Z over a long-term horizon, the findings and implications drawn from this study provide useful information to planners and policymakers on the possible effects of demographic and social changes on building less auto-centric communities with adequate public transport services.

Admittedly, this study has some limitations that are mainly related to the research dataset. Like most of the related literature, I can only interpret the findings of this study as associations rather than causations. Future studies should consider the combination of longitudinal data collection and machine learning algorithms to uncover causality as well as achieve higher prediction accuracy for the behavioral changes across generations. Due to the data limitations, this study cannot precisely control the effects of licensure regulations at each state. Future research could solicit local agencies for state-level licensure regulations and investigate the exact effects on youth licensing and automobility over time.

As the research focus is not to reveal how the built environment influence two generations' travel patterns differently, this study analyzes the publicly available information from the NHTS. Population density, size of MSA, and access to heavy rail status that are hypothesized to be associated with travel outcomes have been taken into consideration. Measuring distinct components of the built environment independently (e.g., land use mix, street networks, and transit service) and exploring their influences on generational differences in youth licensing could be one possible direction for future research. In addition, the NHTS data did not contain adequate stated reference information. To have a better understanding of socio-cultural and geographical differences across population segments, the attitudinal information should be incorporated into the next waves of data collections. This information also helps to explain why young people delay or forgo a driver's license.

Finally, when autonomous vehicles (AVs) become mainstream in the near future, all travelers will relieve the driving task, and therefore having a driver's license may have little impact on types of trips and distances traveled. Future studies in the AVs era could focus on the issues of AVs including legal liability, reaction to driving environment, and performance in a poor weather condition. The results of this study are still meaningful as they reflect which factor may or may not affect generational shifts in preferences and attitudes regarding automobility.

# Appendix

**Table A1. Driving distance and the total number of trips made on the survey day by licensed teens in 2009 and 2017**

|  | 2009 | | 2017 | | 2009 versus 2017 | |
| --- | --- | --- | --- | --- | --- | --- |
|  | Mean | SD | Mean | SD | Changes in % | p value |
| Driving distance (miles) | 21.49 | 32.11 | 24.05 | 46.92 | 11.9% | 0.114 |
| Number of total trips made | 2.69 | 1.83 | 2.25 | 1.78 | -16.5% | 0.000 |
| Number of observations | 3282 | | 3486 | | | |

**Table A2. Tobit models of driving distances by licensed teens (18-20 years old)**

|  | 2009 | | 2017 | | Pooled Model | |
| --- | --- | --- | --- | --- | --- | --- |
|  | Coef. | p value | Coef. | p value | Coef. | p value |
| **Survey year** | | | | | | |
| 2017 (ref: 2009) | | | | | 0.412 | 0.843 |
| **Socio-demographic characteristics** | | | | | | |
| Male (ref: female) | 3.678 | 0.030 | 6.464 | 0.073 | 5.168 | 0.016 |
| Age (ref: 18-year-old) | | | | | | |
|    19-year-old | 4.358 | 0.044 | 1.729 | 0.582 | 3.454 | 0.065 |
|    20-year-old | 7.398 | 0.006 | 3.690 | 0.235 | 6.247 | 0.008 |
| Educational Attainment (ref: less than high school) | | | | | | |
|    High school graduate | 4.986 | 0.059 | 8.304 | 0.006 | 6.376 | 0.002 |
|    Some college or associate degree | 8.198 | 0.002 | 16.419 | 0.000 | 11.619 | 0.000 |
|    Bachelor's and higher | 7.543 | 0.283 | 16.229 | 0.203 | 14.283 | 0.034 |
| Born in US (ref: immigrants) | 8.018 | 0.006 | 6.290 | 0.218 | 6.894 | 0.020 |
| Race of household head (ref: non-Hispanic white) | | | | | | |
|    Hispanic | 1.083 | 0.678 | -2.767 | 0.440 | -0.522 | 0.819 |
|    Non-Hispanic black | -2.602 | 0.381 | -1.985 | 0.635 | -2.036 | 0.432 |
|    Non-Hispanic Asian/Pacific Islander | -1.906 | 0.604 | 7.111 | 0.205 | 3.163 | 0.361 |
|    Other races | 0.802 | 0.870 | 17.353 | 0.337 | 12.120 | 0.315 |
| **Life course events, household incomes, and vehicle ownership** | | | | | | |
| Workers (ref: nonworker) | 8.017 | 0.001 | 22.425 | 0.000 | 15.320 | 0.000 |
| Living independently (ref: living with parents) | 12.217 | 0.217 | 4.388 | 0.468 | 6.424 | 0.217 |
| Percent of family members with a driver's license | -0.458 | 0.904 | 1.363 | 0.814 | 0.956 | 0.801 |
| Household incomes per capita (in 1,000 2017 US$) | 0.121 | 0.001 | 0.021 | 0.592 | 0.070 | 0.008 |
| **Current travel mode choice** | | | | | | |
| Number of walk trips in the past week | -0.110 | 0.496 | -0.567 | 0.095 | -0.401 | 0.056 |
| Number of days used public transit in the past month | -1.096 | 0.000 | -1.753 | 0.008 | -1.446 | 0.000 |
| Frequency of purchased online for delivery in the past month | 0.179 | 0.621 | 0.160 | 0.591 | 0.122 | 0.589 |
| Current bicyclists (people who biked in the past week; ref: non-user) | -6.707 | 0.028 | -9.772 | 0.095 | -8.237 | 0.011 |
| User of rideshare services (e.g., Uber and Lyft) in the past month (ref: non-user) | | | -0.851 | 0.893 | -1.051 | 0.832 |
| User of carshare services (e.g., Zipcar and Car2go) in the past month (ref: non-user) | | | -1.869 | 0.937 | -0.825 | 0.970 |
| **Number of daily trips on the survey day** | | | | | | |
| Social trips | 5.719 | 0.000 | 2.082 | 0.137 | 4.073 | 0.000 |
| Recreation trips | 5.473 | 0.027 | 12.640 | 0.019 | 7.286 | 0.003 |
| Exercise trips | 0.955 | 0.546 | 13.523 | 0.190 | 6.258 | 0.151 |
| Errand trips | 6.692 | 0.000 | 5.728 | 0.001 | 6.403 | 0.000 |
| Education trips | 10.958 | 0.000 | 9.772 | 0.004 | 10.855 | 0.000 |
| Work trips | 8.794 | 0.000 | 7.941 | 0.015 | 8.772 | 0.000 |
| **Residential location choice** | | | | | | |
| Population density (1,000 persons per square mile) | -0.869 | 0.000 | -1.476 | 0.002 | -1.199 | 0.000 |
| Size of metropolitan areas (ref: In an MSA or CMSA of 1,000,000 - 2,999,999 & In an MSA or CMSA of 3 million or more) | | | | | | |
|     Not residing in MSAs | 6.565 | 0.042 | 5.050 | 0.460 | 6.049 | 0.083 |
|     In an MSA or CMSA of population less than 1,000,000 | 3.808 | 0.065 | -1.470 | 0.798 | 1.001 | 0.721 |
| MSA heavy rail status for household (ref: MSA does not have rail, or household not in an MSA) | | | | | | |
|     MSA has rail | 3.484 | 0.139 | 0.040 | 0.996 | 2.042 | 0.546 |
| Constant | -14.864 | 0.384 | -38.222 | 0.070 | -29.477 | 0.042 |
| **Model fitness** | | | | | | |
| Initial log likelihood | | -13745.04 | | -16097.61 | | -30153.05 |
| Final log likelihood | | -13508.42 | | -15974.21 | | -29862.35 |
| Degree of freedoms | | 82 | | 80 | | 84 |
| AIC/BIC | | 27142.67/27636.46 | | 32108.41/32600.93 | | 59892.69/60465.57 |
| Number of observations | | | 3282 | | 3486 | 6768 |

Note: Due to space constraints, we do not report the coefficients for the state control variables; these are available from the authors upon request. MSA = metropolitan statistical area; CMSA = consolidated metropolitan statistical area; AIC = Akaike information criterion; BIC = Bayesian information criterion.

**References**

Allison+ Partners. (2019). HOW THE BIRTH OF MOBILITY CULTURE AND THE RISE OF NEW CONSUMER VALUES WILL REDEFINE OUR JOURNEY FROM HERE TO THERE. Retrieved from https://s3.us-west-2.amazonaws.com/allisonpr.com/201906/7/b/5/20190613154246_5406550/Mobilityreportd3_Print.pdf.

Bates, L. J., Filtness, A., & Watson, B. (2018). Driver education and licensing programs. In Safe mobility: Challenges, methodology and solutions. Emerald Publishing Limited.

Blumenberg, E., Ralph, K., Smart, M., & Taylor, B. D. (2016). Who knows about kids these days? Analyzing the determinants of youth and adult mobility in the US between 1990 and 2009. Transportation Research Part A: Policy and Practice, 93, 39-54.

Blumenberg, E., & Smart, M. (2014). Brother can you spare a ride? Carpooling in immigrant neighbourhoods. Urban Studies, 51(9), 1871-1890.

Brown, R. E., & Handy, S. L. (2015). Factors Associated With High School Students' Delayed Acquisition of a Driver's License: Insights From Three Northern California Schools. Transportation Research Record, 2495(1), 1-13.

Chatman, D. G., & Klein, N. J. (2013). Why do immigrants drive less? Confirmations, complications, and new hypotheses from a qualitative study in New Jersey, USA. Transport policy, 30, 336-344.

Cilliers, E. J. (2017). The challenge of teaching generation Z. PEOPLE: International Journal of Social Sciences, 3(1), 188-198.

Delbosc, A., & Currie, G. (2013). Causes of youth licensing decline: a synthesis of evidence. Transport Reviews, 33(3), 271-290.

Delbosc, A., & Currie, G. (2014). Changing demographics and young adult driver license decline in Melbourne, Australia (1994–2009). Transportation, 41(3), 529-542.

Delbosc, A. (2017). Delay or forgo? A closer look at youth driver licensing trends in the United States and Australia. Transportation, 44(5), 919-926.

Dimock, M. (2019). Defining generations: Where Millennials end and Generation Z begins. Retrieved from: https://www.pewresearch.org/fact-tank/2019/01/17/where-millennials-end-and-generation-z-begins/

Erlbaum, N. (2005). Assessment of 2001 New York State NHTS Add-On Data Using Empirical and Auditable Data Sources. Journal of Transportation and Statistics, 8(3), 1.

Federal Highway Administration, (2004). 2001 national household travel survey user's guide. Retrieved from http://nhts.ornl.gov/2001/usersguide/UsersGuide.pdf.

Federal Highway Administration, (2011). 2009 national household travel survey user's guide. Retrieved from https://nhts.ornl.gov/2009/pub/UsersGuideV2.pdf

Federal Highway Administration, (2018). 2017 national household travel survey user's guide. Retrieved from https://nhts.ornl.gov/assets/2017UsersGuide.pdf

Freedman, D., Pisani, R., Purves, R. and Adhikari, A. (1991) Statistics, 2nd edn. New York: Norton.

Garikapati, V. M., Pendyala, R. M., Morris, E. A., Mokhtarian, P. L., & McDonald, N. (2016). Activity patterns, time use, and travel of millennials: a generation in transition? Transport Reviews, 36(5), 558-584.

Habib, K. N. (2018). Modelling the choice and timing of acquiring a driver's license: Revelations from a hazard model applied to the University students in Toronto. Transportation Research Part A: Policy and Practice, 118, 374-386.

Handy, S., Wang, A., Jacobson, E., & Thigpen, C. (2021). What do teenagers think about driving? Insights from a bicycling-oriented community in the auto-dependent United States. Transportation research interdisciplinary perspectives, 11, 100422.

Hirschberg, J., & Lye, J. (2020). Impacts of graduated driver licensing regulations. Accident Analysis & Prevention, 139, 105485.

Hjorthol, R. (2016). Decreasing popularity of the car? Changes in driving licence and access to a car among young adults over a 25-year period in Norway. Journal of Transport Geography, 51, 140-146.

Jamme, H. T., Rodriguez, J., Bahl, D., & Banerjee, T. (2019). A Twenty-Five-Year Biography of the TOD Concept: From Design to Policy, Planning, and Implementation. Journal of Planning Education and Research, 39(4), 409-428.

Katz, R., Ogilvie, S., Shaw, J. and Linda Woodhead, L. (2019), "Understanding the iGeneration. Center for advanced study in the behavioral sciences", available at: https://casbs.stanford.edu/projects/projects/understanding-igeneration

Le Vine, S., Latinopoulos, C., & Polak, J. (2014). What is the relationship between online activity and driving-license-holding amongst young adults?. Transportation, 41(5), 1071-1098.

Lee, Y., & Circella, G. (2019). ICT, millennials' lifestyles and travel choices. In Advances in transport policy and planning (Vol. 3, pp. 107-141). Academic Press.

Lee, Y., Circella, G., Mokhtarian, P. L., & Guhathakurta, S. (2020). Are millennials more multimodal? A latent-class cluster analysis with attitudes and preferences among millennial and Generation X commuters in California. Transportation, 47(5), 2505-2528.

Lee, S., Smart, M. J., & Golub, A. (2021). Difference in travel behavior between immigrants in the US and US born residents: the immigrant effect for car-sharing, ride-sharing, and bike-sharing services. Transportation research interdisciplinary perspectives, 9, 100296.

Lee, H. (2020). Are millennials coming to town? Residential location choice of young adults. Urban affairs review, 56(2), 565-604.

Luttrell, R., & McGrath, K. (2021). Gen Z: The superhero generation. Rowman & Littlefield Publishers.

Mannheim, Karl. 1952. "The Problem of Generations." In Essays on the Sociology of Knowledge, edited by Karl Mannheim, 276–322. London: Routledge & Kegan Paul.

McDonald, N., & Trowbridge, M. (2009). Does the built environment affect when American teens become drivers? Evidence from the 2001 National Household Travel Survey. Journal of safety research, 40(3), 177-183.

McDonald, N. C. (2015). Are millennials really the "go-nowhere" generation?. Journal of the American Planning Association, 81(2), 90-103.

Olsson, L. E., Friman, M., Lättman, K., & Fujii, S. (2020). Travel and life satisfaction-From Gen Z to the silent generation. Journal of Transport & Health, 18, 100894.

Parker, K., & Igielnik, R. (2020). On the Cusp of Adulthood and Facing an Uncertain Future: What We Know About Gen Z So Far. Retrieved from https://www.pewresearch.org/social-trends/2020/05/14/on-the-cusp-of-adulthood-and-facing-an-uncertain-future-what-we-know-about-gen-z-so-far-2/

Pew Research Center. 2018. http://www.pewresearch.org/topics/ millennials/

Pucher, J., & Renne, J. L. (2003). Socioeconomics of urban travel. Evidence from the 2001 NHTS. Federal Highway Administration, (2004). 2001 national household travel survey user's guide. Retrieved from http://nhts.ornl.gov/2001/usersguide/UsersGuide.pdf.

Rérat, P. (2021). A decline in youth licensing: a simple delay or the decreasing popularity of automobility?. Applied Mobilities, 6(1), 71-91.

Rue, P. (2018). Make way, millennials, here comes Gen Z. About Campus, 23(3), 5-12.

Ryder, N. B. (1985). The cohort as a concept in the study of social change. In Cohort analysis in social research (pp. 9-44). Springer, New York, NY.

Shin, E. J. (2017). Unraveling the effects of residence in an ethnic enclave on immigrants' travel mode choices. Journal of Planning Education and Research, 37(4), 425-443.

Shirgaokar, M., & Nobler, E. (2021). Differences in daily trips between immigrants and US-born individuals: Implications for social integration. Transport policy, 105, 103-114.

Smart, M. J., & Klein, N. J. (2018). Remembrance of cars and buses past: How prior life experiences influence travel. Journal of Planning Education and Research, 38(2), 139-151.

Statista (2021). US population by generation 2020. Retrieved from https://www.statista.com/statistics/797321/us-population-by-generation/

Strauss, William, and Neill Howe. 1997. The Fourth Turning: An American Prophecy. Portland: Broadway Books

Tefft, B. C., Williams, A. F., & Grabowski, J. G. (2014). Driver licensing and reasons for delaying licensure among young adults ages 18-20, United States, 2012. Injury epidemiology, 1(1), 1-8.

Thach, L., Riewe, S., & Camillo, A. (2020). Generational cohort theory and wine: analyzing how gen Z differs from other American wine consuming generations. International Journal of Wine Business Research.

Thigpen, C., & Handy, S. (2018). Driver's licensing delay: A retrospective case study of the impact of attitudes, parental and social influences, and intergenerational differences. Transportation research part A: policy and practice, 111, 24-40.

Twenge, J. M. (2017). iGen: Why today's super-connected kids are growing up less rebellious, more tolerant, less happy--and completely unprepared for adulthood--and what that means for the rest of us. Simon and Schuster.

US Census Bureau. (2020). 2020 American Community Survey (ACS) 5-year estimates.

van der Waard, J., Jorritsma, P., & Immers, B. (2013). New drivers in mobility; what moves the Dutch in 2012?. Transport Reviews, 33(3), 343-359.

Wang, K., & Akar, G. (2020). Will Millennials Drive Less as the Economy Recovers: A Postrecession Analysis of Automobile Travel Patterns. Journal of Planning Education and Research, 0739456X20911705.

Wang, X. (2019). Has the relationship between urban and suburban automobile travel changed across generations? Comparing Millennials and Generation Xers in the United States. Transportation Research Part A: Policy and Practice, 129, 107-122.

Williams, A.F., 2011. Teenagers' licensing decisions and their views of licensing policies: a national survey. Traffic Inj. Prev. 12, 312–319.

Williams, A. F., McCartt, A. T., & Sims, L. B. (2016). History and current status of state graduated driver licensing (GDL) laws in the United States. Journal of safety research, 56, 9-15.

Winship, C., & Radbill, L. (1994). Sampling weights and regression analysis. Sociological Methods & Research, 23(2), 230-257.

Witmer, D. (2019, July 18). How Old Does Your Teen Need to Be to Legally Drive? Retrieved from https://www.verywellfamily.com/driving-age-by-state-2611172

Wu, C., Le Vine, S., & Sivakumar, A. (2021). Exploratory analysis of young adults' trajectories through the UK driving license acquisition process. Traffic injury prevention, 22(1), 37-42.